\documentclass[preprint,aps]{revtex4-1}

\usepackage[dvips]{graphicx}
\usepackage{amssymb,amsfonts,amsmath}
\usepackage[usenames]{color}

\begin{document}

\title{A comparative study on bulk and nanoconfined water by time-resolved optical Kerr effect spectroscopy}

\author{Andrea Taschin$^{1}$, Paolo Bartolini$^{1}$, Agnese Marcelli$^{1}$, Roberto Righini $^{1,2}$, and Renato Torre$^{1,3}$}

\affiliation{
$^1$European lab. for Non-Linear Spectroscopy (LENS), Univ. di Firenze, via N. Carrara 1, I-50019 Sesto Fiorentino, Firenze, Italy.\\
$^2$Dip. di Chimica, Univ. di Firenze, via Della Lastruccia 13, I-50019 Sesto Fiorentino, Firenze, Italy.\\
$^3$Dip. di Fisica e Astronomia, Univ. di Firenze, via Sansone 1, I-50019 Sesto Fiorentino, Firenze, Italy.}

\date{\today}


\begin{abstract} 
{The low frequency ($\nu <$ 500~cm$^{-1}$) vibrational spectra of hydrated porous silica are specifically sensitive to the hydrogen bond interactions and provides a wealth of information on the structural and dynamical properties of the water contained in the pores of the matrix. We investigate systematically this spectral region of Vycor porous silica (pore size $\simeq$ 4 nm) for a series of samples at different levels of hydration, from the dry matrix to completely filled pores. The spectra are obtained as the Fourier transforms of time-resolved heterodyne detected optical Kerr effect (HD-OKE) measurements. The comparison of these spectra with that of bulk water allows us to extract and analyze separately the spectral contributions of the first and second hydration layers, as well as that of bulk-like inner water. We conclude that the extra water entering the pores above $\approx$ 10~\% water/silica weight ratio behaves very similarly to bulk water. At lower levels of hydration, corresponding to two complete superficial water layers or less, the H-bond bending and stretching bands, characteristic of the tetrahedral coordination of water in the bulk phase, progressively disappear: clearly in these conditions the H-bond connectivity is very different from that of liquid water. A similar behavior is observed for the structural relaxation times, measured from the decay of the time-dependent HD-OKE signal. The value for the inner water is very similar to that of the bulk liquid; that of the first two water layers is definitely longer by a factor $\sim$ 4. These findings should be carefully taken into account when employing pore confinement to extend towards lower temperatures the accessible temperature range of supercooled water.}
\end{abstract}
\maketitle

\section{Introduction}

The surface interactions between a liquid and a solid produce local modification on both material properties. In particular, the liquid layers at the interface show structural and dynamic alterations that turn into non-trivial modification of the fundamental chemical-physical properties of the liquid. Thus, in case of complete spatial confinement the behaviour of the liquid is determined by its interfacial properties\cite{brovchenko_08}. 

The investigation and study of confined liquids is relevant to a broad variety of scientific topics, spanning from biology to geophysics. Recently, the studies aimed to understand water anomalies have boosted a large interest on the proprieties of water confined in silica nanopores. It is well known that a number of chemical-physical properties of bulk water differ from those of the other liquids\cite{debenedetti_03,debenedetti_03b}. In particular the temperature dependence of several experimental observables have unexpected and counter-intuitive behaviours. For example, some thermodynamic quantities (e.g. isothermal compressibility, isobaric heat capacity) show a sharp increase upon cooling, as well some dynamic features (e.g. viscosity and structural relaxation time) present a critical slowing down when temperature is lowered. Many experimental and simulation works suggested the existence of a water singularity temperature, T$_s$ around 223-228 K at atmospheric pressure, that should reflect the existence of a critical process taking place in the supercooled water phase (i.e.  the metastable non-equilibrium phase where water enters when is cooled below the melting point and crystallization does not take place). The nature of this critical phenomenon is still largely debated: some authors postulate the existence of a liquid-liquid phase transition~\cite{poole_92,poole_13}, another approach sees the water anomalies as non-equilibrium phenomena precursory of the liquid-crystal transition~\cite{limmer_11}. Moreover, the relevance of ice spontaneous nucleation has been recognized~\cite{moore_09,moore_11}.
A relevant experimental drawback is that the critical evidences are expected to occur at temperature/pressure values not directly accessible in bulk samples, and this prevents a direct experimental solution of the problem; homogeneous nucleation fixes the lowest reachable temperature at about 231 K for bulk water. Actually, in a macroscopic bulk sample the crystallization phenomena and the experimental difficulties limit the minimum temperatures around 243 K. 

Nanoconfinement has the advantage of preventing water freezing and of enabling the experimental investigation of the supercooled phase down to very low temperatures.  
Nevertheless, the water crystallization doesn't take place only if the confinement is very tight, reaching the nanometric length scale. Already the confinement of water in silica nanopores of 10 nm diameter produces a lowering of the crystallization temperature of about 15 degrees. In order to maintain the water liquid approaching T$_s$ a confinement of about 1.5 nm is required, in pore of diameter  $\leq$ 1 nm water remains in a liquid-like phase also below that singularity temperature.

Here comes the basic question: to what extent the interactions with the pore surfaces modify the water properties ?

The debate on this issue is open\cite{cucini_10,taschin_10,cucini_10b,gallo_10,molinero_10,capaccioli_11,lecaer_11,limei_11,leporini_12,limmer_12,milischuk_12,alabarse_12,giovambattista_12}. All the researchers agree that some structural and dynamic properties of water at the silica interface are modified from those of the bulk liquid, but there is no agreement about if these modifications  involve also the water fraction situated in the inner part of the pore, not in direct contact with the surface.

In order to give a contribution to the understanding of this problem, we undertook a comparative experimental investigations of bulk and nanoconfined water by means of time-resolved optical Kerr effect spectroscopy. 

\section{Optical Kerr effect experiments}

In an optical Kerr effect (OKE) experiment~\cite{taschin_13,torre_08,hunt_07,mcmorrow_88} a linear-polarized short laser pulse (the pump) induces a transient birefringence in an optically transparent medium. The induced birefringence is measured by monitoring the polarization changes produced in second laser pulse (the probe) which is spatially superimposed with the pump pulse into the sample. The time behaviour of the induced birefringence is reconstructed by changing the time delay between the pump and probe.

The resulting signal reflects the relaxation and vibrational response of the molecules in the sample. The method allows measuring the fast relaxation processes and the low frequency vibrational correlation functions of the system by  employing femtosecond laser pulses. OKE experiment is the time-domain counterpart of depolarized light scattering technique~\cite{taschin_13,torre_08,hunt_07}. In our case, the laser system is a self-mode-locked Ti:sapphire laser producing pulse of 20 fs duration with energy of 3 nJ. A detailed description of the optical set-up and of the experimental apparatus is reported in references\cite{bartolini_09,taschin_13}. Our set-up enables measurements in a wide time window (from tens of femtoseconds to hundreds of picoseconds), making OKE a very flexible technique, able of revealing very different dynamic regimes, from that typical of simple liquids to that characterizing supercooled liquids and glass formers~\cite{torre_04,bartolini_99,torre_98,torre_00}. This is obtained by implementing a peculiar configuration for the heterodyne detection~\cite{bartolini_09,hunt_07,giraud_03}, with circularly polarized probe beam and differential acquisition of two opposite-phase signals on a balanced double photodiode. Thus, the measured OKE signal is automatically heterodyned and free from possible spurious signals. Finally, the photodiode output signal is processed by a lock-in amplifier phase locked to the reference frequency at which the pump beam is chopped.

With optical heterodyne detection, the OKE signal is directly proportional to the material response function, $R(t)$, convoluted with the instrumental function, $G(t)$. The response $R(t)$ is directly connected to the time derivative of the time-dependent correlation function of the dielectric susceptibility \cite{taschin_13,torre_08,hunt_07}:
\begin{equation}
R(t)\propto \frac{\partial}{\partial t}\langle\chi(t)\chi(0)\rangle
\label{signaleq}
\end{equation}
The Fourier transform of $R(t)$ corresponds to the frequency-dependent response measured in depolarized light scattering (DLS) experiments~\cite{taschin_13,torre_08,hunt_07}. 

In order to get access to the real OKE response, the knowledge of the correct instrumental function $G(t)$ is fundamental, especially for the short delay time range. This is particularly critical for weak signals characterized by complex relaxation dynamics, as in the case of bulk  water or confined water.
In short, the critical point in the acquisition of the instrumental function is that any minor, apparently negligible, adjustment of the optical set-up when passing from recording the instrumental function to measuring the water signal, prevents the extraction of the real instrumental function. The resulting inaccurate deconvolution of the water signal degrades the quality of the overall fitting. To minimize these effects, $G(t)$ was obtained in our experiment by measuring the OKE signal of a CaF$_2$ plate of the same thickness of Vycor sample placed side by side in the same cell. The switching from the water measurement configuration to the instrumental one was achieved by simply translating the cell perpendicularly to the optical axis of the experiment without even touching the rest of optical set-up\cite{taschin_13}.

Our samples are Vycor slabs (code 7930 by Corning Company) of 8x8x2 mm$^{3}$ dimensions, with porosity 28\% of volume, internal surface area S=250  m$^{2}$/g, average pore diameter of 4 nm and density of 1.5 g/cm$^{3}$ (in dry condition), according to the manufacturer technical data sheet. The samples were cleaned with 35\% hydrogen peroxide solution (heating up to 90 $^\circ$C for 2 hours) and washed in distilled water. They were then stored in P$_{2}$O$_{5}$ (phosphoric anhydride) until usage. All the measurements have been carried out at room temperature.

The HD-OKE experiments require a very good optical quality of the samples investigated. Vycor porous glass is among the few porous glasses having a solid structure that enables to prepare macroscopic sample with polished surfaces of optical quality. The surfaces of our samples have been optically polished by an abrasive paste.

In spite of the good optical quality of the surfaces, the Vycor samples have, especially at low hydration level, an intrinsic scattering ability probably due to the presence of heterogeneities whose size is comparable to the laser wavelength. Part of scattered pump light propagates collinearly to the probe and interferes with it on the photodetector, giving rise to an unwanted signal. This spurious contribution appears only during the time superposition of the two pulses and degrades the quality of the OKE data at very short delay time. This disturbing signal was removed by inserting a piezo-driven vibrating mirror in the optical path of the pump arm. We set the amplitude of the mirror stroke and the frequency of the sinusoidal driving signal in a way that the induced frequency modulation transfers the spectral content of the spurious signal out of the acceptance band of the lock-in amplifier (the bandwidth is determined by the chopper frequency and by the bandwidth of the notch-filter at the lock-in amplifier input). This device allowed us to almost completely remove the unwanted contribution.

\section{Results on bulk liquid water}

The hydrogen bond network largely determines the intra- and inter-molecular vibrational spectrum of liquid water; thus the experimental investigation of the spectral features gives precious information on the local H-bond network structures. The investigation of the high vibrational spectrum, 1000-4000 cm$^{-1}$, gives access to the intramolecular vibrational dynamics. The low frequency spectrum of water, $\nu < $ 1000 cm$^{-1}$, reflects its intermolecular dynamics and so it is particularly sensitive to local molecular structures. This frequency range of the water spectrum has been investigated in several light scattering studies, see for example\cite{krishna_83,aliotta_86,walrafen_86,rousset_90,mizoguchi_92,sokolov_95,water_web_site}. 
Two main broad peaks are observed around 50 and 175~cm$^{-1}$ at room temperature, generally attributed to the hydrogen bond ``bending'' and``stretching'' vibrations, respectively. The spectra show also a very low frequency wing, $\nu<$ 20~cm$^{-1}$, which has been attributed to ``relaxation'' processes\cite{aliotta_86}. The light scattering data were measured at relatively high temperatures and have rather poor signal-to-noise ratio, due to the very weak water signal: these drawbacks did not allow identifying all the possible vibrational modes possibly contributing to the water spectrum in this region. 
Time-resolved non-linear spectroscopy has been shown in recent years to be a very useful tool for the investigation of complex liquid dynamics\cite{torre_08,hunt_07}, and in particular of liquid water\cite{castner_95,palese_96,winkler_02,torre_04,ratajska_06,taschin_13}.
\begin{figure}
\begin{center}
\includegraphics[scale=.8]{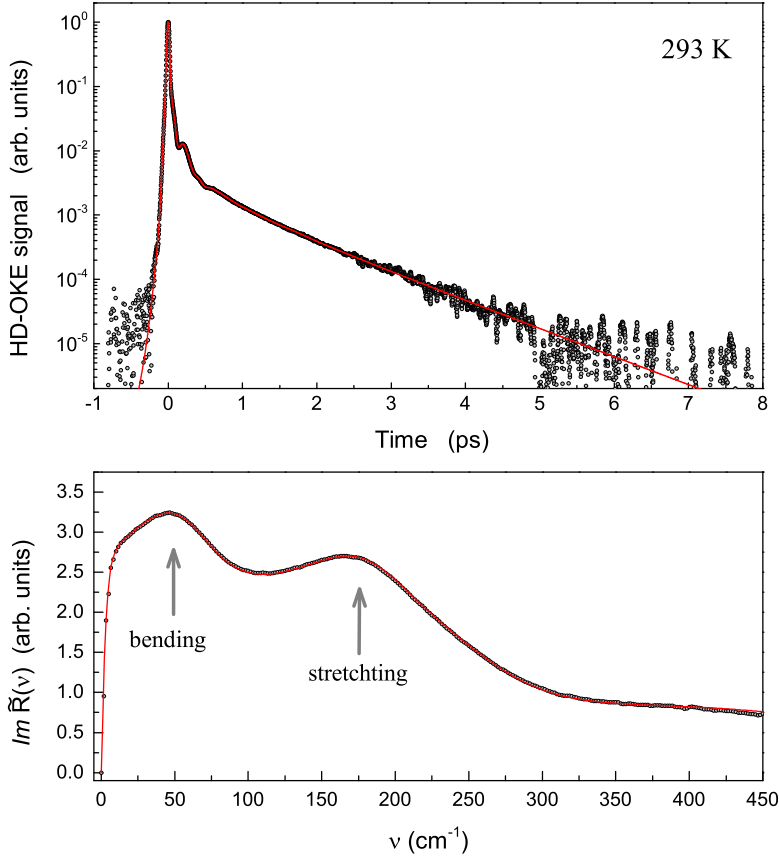}
\caption{We report here a HD-OKE data of bulk water at 293 K. In the upper panel, the experimental data are reported in the time scale with the best fit result (red-line). The data are reported in a log-linear scale. In the lower panel, we show the HD-OKE response function Fourier transform in the frequency domain. The water spectrum displays the two main vibrational bands: bending and stretching located around 50 and 175 cm$^{-1}$, respectively.}
\label{fig_1}
\end{center}
\end{figure}
Recent heterodyne-detected optical Kerr effect (HD-OKE) investigations of liquid water~\cite{torre_04,taschin_13} at temperature below the melting point (i.e. in the supercooled phase) are reporting interesting evidence of critical phenomena.

In fig.~\ref{fig_1} we report the HD-OKE data of liquid water at 293 K: the data clearly show the signature of fast vibrational dynamics at short times, extending up to 1 ps, followed by a slower monotonic relaxation. The initial oscillatory component provides information about the intermolecular hydrogen-bond dynamics, and can be directly compared to the low-frequency Raman spectrum\cite{castner_95}. At longer times the data display a monotonic decay due to structural relaxation. The relaxation features presented by water are very similar to those of the glass-former liquids\cite{torre_00}, and the slow dynamics is found in agreement with the main mode-coupling theory predictions\cite{torre_04}.
For the analysis of the whole time-dependent correlation function measured in optical experiments a model is required that enables an operative parametrization of the vibrational modes present in the water dynamics. This turns out to be a complex task, and different attempts have been pursued\cite{taschin_13,palese_96,winkler_02,ratajska_06}. We decided to use a relative simple approach, that allows the comparative analysis of HD-OKE data obtained both in bulk and nanoconfined water. 

The data are fitted in the time scale according to the following expressions:
\begin{eqnarray}
S(t)=\int \left[ k\delta(t-t')+R(t-t') \right]  G(t') dt';\label{signal}
\\
R(t)=B\frac{d}{dt} exp\left[-\left(\frac{t}{\tau}\right)^{\beta}\right]+\sum_i C_i exp\left(-\gamma_i^2 t^2\right)\sin\left(\omega_i t\right)
\label{response}	
\end{eqnarray}
Eq.~\ref{signal} describes the convolution of the response function, $R(t)$, with the instrumental function, $G(t)$, where the $\delta$-function reproduces the instantaneous electronic response\cite{hellwarth_77}. Eq.~\ref{response} gives the response function simulating the material dynamics. The liquid water dynamics is described by the sum of a relaxation function in the form of a stretched exponential\cite{torre_04}, and of a few damped oscillators (DO). We are aware that this is a simple fitting function that turns to be inappropriate to reproduce the supercooled water HD-OKE data at low temperatures\cite{taschin_13}, nevertheless it is sufficient to describe the liquid water at relative high temperature or/and in nanoconfinement. The fitting parameters of the model are: the structural relaxation time $\tau$, the stretching factor $\beta$,  the frequency $\omega_i$ and the damping constants $\gamma_i$ of the DOs.

In the upper panel of fig.~\ref{fig_1} we show the fit, obtained with the eq.s~\ref{signal} and \ref{response}, of the measured HD-OKE signal: the fitting function is able to reproduce correctly the experimental data over the whole time window. The parameters used to fit the data are; structural relaxation: $\tau$=0.35 ps and $\beta$=0.6; vibrational dynamics: the bending mode at about 50 cm$^{-1}$ is described by 2 DOs ($\omega_1$= 47 cm$^{-1}$, $\omega_2$=100 cm$^{-1}$) and the stretching mode around 175 cm$^{-1}$ needs 2 DOs ($\omega_3$=173 cm$^{-1}$ and $\omega_4$=238 cm$^{-1}$) to be properly reproduced. 

The HD-OKE data, collected in time domain, can be Fourier transformed in the frequency domain to yield the corresponding spectra. This enables a more immediate visualization of the different vibrational modes present in the water dynamics and a direct comparison with the depolarized light scattering experiments\cite{taschin_13,torre_08,hunt_07}. In order to get the spectra of the HD-OKE response function, we need to: i) Fourier transform the measured data, ii) deconvolute them from the instrumental response and iii) retain the imaginary part of it, i.e. $Im[\tilde{R}(\nu)] \propto Im\lbrace FT\left[ S(t) \right] /FT\left[ G(t)\right] \rbrace$. The details about these procedures are reported in Taschin et al.\cite{taschin_13}.  

The $Im[\tilde{R}(\nu)]$ obtained from the HD-OKE data and the corresponding fits are reported in lower panel of fig.~\ref{fig_1}. The frequency response shows clearly the main vibrational features present in the liquid water spectrum: the two bands around 50 and 175~cm$^{-1}$, the ``bending" and ``stretching" modes. These vibrations are characteristic of the intermolecular dynamics of the first neighbour molecular cage\cite{skaf_05,desantis_04,padro_04}, that for liquid water are largely determined by the hydrogen bonds. The shoulder appearing at very low frequency, $\nu \lesssim$ 10 cm$^{-1}$, is due to the relaxation processes. As expected, this features is hardly detectable in the frequency domain, whereas the vibrational modes are clearly visible. 

\section{Results on nanoconfined water}

\subsection{Hydration control and FT-IR spectroscopy}

We studied Vycor glasses characterized by a variable water filling, from ``dry'' to ``fully hydrated''. The hydrophilic properties of porous silica samples depend on the degree of surface hydroxylation, i.e. on the amount of Si-OH groups. This quantity determines the potential hydration rate of the surface and may influence also the structure of the first water layer, in which the molecules interact with the silanol groups. In the literature, the silanols are often indicated as chemical (or chemisorbed) water, as they are the result of dissociative chemisorption of water molecules onto the silica surface, while the non-dissociated molecular water interacting with the surface is referred to as physical (or physisorbed) water\cite{Zhuravlev2000,Burneau1997}. 

In spite of the amount of work done on this subject, water filling in silica nanopores is still a debated issue\cite{gallo_02,molinero_10}. Basically, two different hypothesis have been formulated. According to the first model, water fills the pore following a layer by layer process until the full hydration level is reached\cite{gallo_02}. A second different filling process has been described in a recent study\cite{molinero_10}: at low hydration level water fills the pore forming the first layer, as in the previous model, but at higher hydration water starts to form liquid plugs connecting the pore surfaces that coexists with the surface-adsorbed water.

Fourier Transform Infrared (FT-IR) spectroscopy is a valid tool  to qualitatively and quantitatively characterize the hydration process of porous glass. Spectra were recorded by an ALPHA FT-IR spectrometer (Bruker Optics) in the range of 4200-5500 cm$^{-1}$ , where two absorption bands can be  well-distinguished at $\sim$ 4550 and $\sim$ 5260 cm$^{-1}$ arising from combination  of stretching and bending modes of silanol and water, respectively\cite{Gallas1989}.

Dry sample were obtained by heating Vycor at 400 $^\circ$C for 10 h, while samples at different hydration level have been prepared by exposing dry Vycor slabs at water atmosphere for 12 h in closed vials, previously purged with nitrogen. Samples with different amount of water filled nanopores were obtained \textit{via} a vapor phase exothermic transfer from the bulk\cite{Tombari2005}. The water concentration in each sample was adjusted adding different volume of bulk water into the vial containing Vycor by means of microsyringe. After this procedure, all the samples were weighted for checking the water content and immediately closed into a hermetic cell for spectroscopic measurements. No significant variation of the sample mass was found after the OKE measurements.. 

\begin{figure}
        \begin{center}
       \includegraphics[scale=.7]{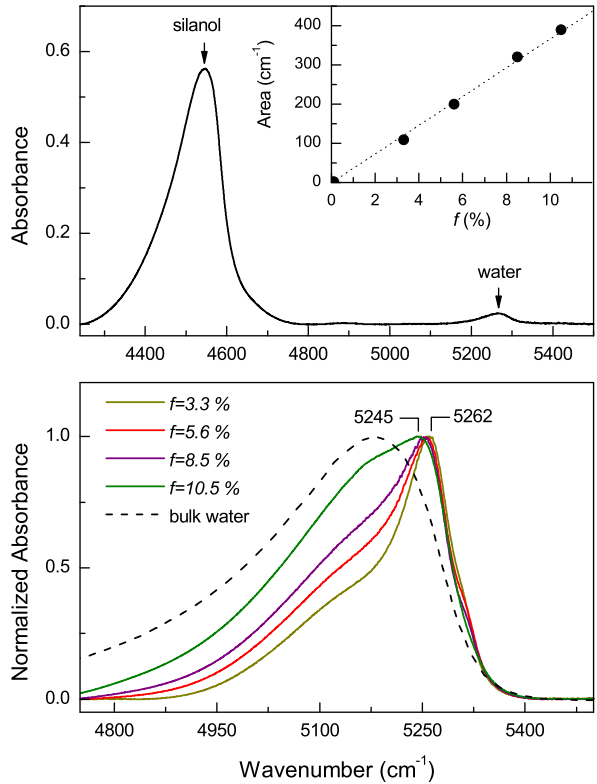}
              \caption{
Upper panel: FTIR spectrum of Vycor  after heating the sample up to 400 $^\circ$C for 10 h. The absorption bands of silanol and water are indicated. In the \textit{inset}, we show the variation of the area value of the water band at $\sim$ 5260 cm$^{-1}$ as a function of hydration level.  This is measured by the filling fraction, \textit{f}, defined as the ratio between water and silica weights.\\
Lower panel: Absorption band of water on partially filled Vycor glass at variable hydration level, \textit{f}. The absorption band recorded in bulk water\citep{water_web_site} is also reported.}
        \label{fig_2}
        \end{center}
   \end{figure}

The hydration level can be quantify using the \textit{Filling Fraction} parameter: $f$=H$_2$O (g) / Vycor (g). This is defined as the ratio between the weight of water contained into Vycor glass and the weight of the borosilicate glass composing the Vycor matrix. Even if the heating temperature of 400 $^\circ$C is over the temperature needed to remove the absorbed physical water, some re-absorption occurs during the cooling process and the weighting procedure. Indeed, in the spectrum of Vycor sample recorded after heating (upper panel of fig.~\ref{fig_2}) the absorption band of water at 5260 cm$^{-1}$ has very low intensity. From the area of this band we can estimate the actual amount of physical water left in the Vycor glass. From the molar integrated absorption coefficient, $\varepsilon_{H_{2}0}=$ 0.22 (cm $\mu$mol$^{-1}$)\cite{Gallas1989}, we estimated that the concentration of water in the sample was equal to $f$=0.1~\%. Thus we obtained a good measure of the mass of the completely dehydrated slab.

In summary, we prepared samples at 6 hydration levels, whose water content was accurately determined from their weights: 1) $f$= 0.10 $\pm$ 0.02 \%, 2) $f$=3.3 $\pm$ 0.2 \%, 3) $f$=5.6 $\pm$ 0.3 \%, 4)  $f$=8.5 $\pm$ 0.4 \%, 5) $f$=10.5 $\pm$ 0.5 \% and 6) $f$=24.3 $\pm$ 0.9 \%.  
We can consider the sample (1) as substantially ``dry'' Vycor, in fact at this hydration the number of water molecules inside the pores are much less than the number of silanol groups present on the pore surface. On account of the porous glass properties, the hydration of 5.6~\% (sample (3)) insures enough water inside pores to generate the ``first layer'', i.e. the full coverage of the pore surface with a mono-molecular water layer. Sample (5) (10.5~\% hydration) contains about ``two layers'', whereas the sample (6) (24.3~\% hydration) is in the ``full hydration'' condition\cite{Tombari2005}. We verified that for the partially hydrated samples the integrated absorption area linearly increases with the absorbed water concentration (see the inset of fig.~\ref{fig_2}).   
The spectrum of ``dry" Vycor also shows prominent band at 4550 cm$^{-1}$ due to silanol absorption. Its integrated area value allowed us to evaluate the OH surface concentration, in accordance with the following relationship:
\begin{eqnarray}
n_{OH}=\frac{Area}{\varepsilon_{OH}\cdot{l}}\cdot\frac{1}{\rho}\cdot\frac{1}{S}\cdot{10^{-24}}\cdot{N_{A}}
\label{OH concentration}	
\end{eqnarray}
where $n_{OH}$ indicates the number of OH groups per nm$^{2}$, $\varepsilon_{OH}$ is the  molar integrated absorption coefficient (= 0,16 cm $\mu$mol$^{-1}$)\cite{Gallas1989}, $l$, $\rho$ and $S$  are the thickness (0.2 cm), the density (1.5 g/cm$^{3}$) and the internal surface  area (250 m$^{2}$/g), respectively,  and ${N_{A}}$ the Avogadro's number. We obtained a value of  5 OH per nm$^{2}$, in agreement with the results obtained by Zhuravlev \citep{Zhuravlev2000} on silica surface of amorphous materials. The determination based on the integrated absorption coefficient of silanol band has the advantage to do not significantly vary in the presence of H-bond with surrounding water molecules.  

Moreover, the comparison between the absorption spectra of physical water, reported in the lower panel of fig. \ref{fig_2}, clearly indicates the structure of water drastically changes as a function of water content. The band maximum shifts to the red with increasing of water  concentration, $f$, while a broad absorption grows in below 5200 cm$^{-1}$ gradually approaching the spectral profile of bulk water. The sharp peak at $\sim$ 5260 cm$^{-1}$ is reasonably assigned to vibrations of  monomeric water, the  H-bonding interactions can reasonably lead to the broad red-shifted absorption. 
The latter band is present even at the lowest water concentration ($<$ 5~\%), showing that water aggregation sets up before full surface coverage is attained.

\subsection{HD-OKE data}

\begin{figure}
\centering
\includegraphics[scale=.8]{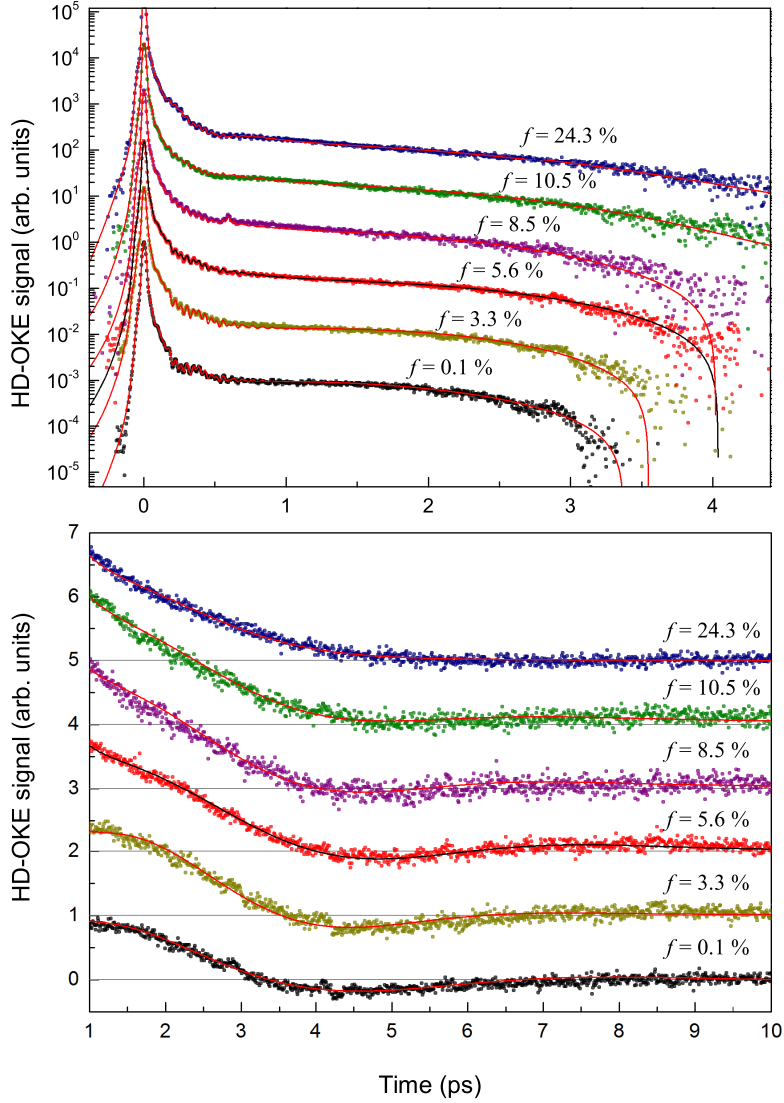}
\caption{
HD-OKE data of nanoconfined water at 293 K and at variable hydration levels, from dry condition, $f$=0.1~\%, to full hydration, $f$=24.3~\%. Full circles: experimental data;  continuous line: best fit results. In the upper panel the data are reported in a log-linear plot. In the lower panel the signal decays at longer times are shown in linear-linear plot. The data have been vertically shifted to make all kinetics clearly visible.
}
\label{fig_3}
\end{figure}

We measured the HD-OKE response at room temperature, 293 K, of the six Vycor samples at different hydration levels section ($f$= 0.1~\%, 3.3~\%, 5.6~\%, 8.5~\%, 10.5~\% and 24.3~\%) described in the previous section. The experimental results are collected in fig.~\ref{fig_3}. The upper panel shows in a log-linear plot the data in the short time range; the results for longer delay times are shown in the lower panel in a linear-linear plot. Just a simple look to the data reveals several differences between the water-in-Vycor and that of bulk water shown in fig. \ref{fig_1}. The data measured at different hydration levels enables to disentangle the nanoconfined water signal from the contribution of the dry Vycor matrix. As shown in fig.~\ref{fig_3}, the signal on dry Vycor (sample (1), $f$=0.1~\%), displays a relatively slow oscillation extending in the picosecond time-scale. According to our fit, it corresponds to an oscillatory mode characterized by a frequency of $\nu \simeq 5$ cm$^{-1}$ and it likely due to an acoustic-like vibration localized on the pore surface\cite{vacher_90}. The presence of liquid water adds a monotonic decay that becomes the dominant feature at full hydration; this contribution is attributed to the relaxation processes of nanoconfined water. 

The fast vibrational components visible in the upper panel of fig.~\ref{fig_3} are rather complex: their characteristic features become clearer when transformed into the frequency domain.
In order to fit the HD-OKE signal we use the equation introduced in the previous section, see eq.s~\ref{signal} and \ref{response}. The response function must be completed to include the dry Vycor matrix response, $R_{dry}(t)$. So the complete response becomes:
\begin{eqnarray}
R(t)=AR_{dry}(t)+B\frac{d}{dt} exp\left[-\left(\frac{t}{\tau}\right)^{\beta}\right]+\sum_i C_i exp\left(-\gamma_i^2 t^2\right)\sin\left(\omega_i t\right)
\label{responsetot}	
\end{eqnarray}
We simulated the $R_{dry}(t)$ response as the sum of a series of damped oscillators and exponential functions, whose characteristic parameters were determined by fitting of the HD-OKE signal of sample (1)($f=0.1$~\%).  

The best fit procedure enables to get the structural relaxation times  of nanoconfined water, these are reported in tab.~\ref{slowdata}. As expected, the amplitude of the structural relaxation component decrease with the hydration level, so that at $f$=3.3~\% it becomes no more measurable. 
\begin{table}[h]
\small
\centering
  \caption{
  Structural relaxation parameters obtained by the best fitting of HD-OKE data. $f$ is the filling fraction, $\tau$ the relaxation time and $\beta$ the stretching parameter.  
  }
  \label{slowdata}
  \begin{tabular}{c c c c} 
    \hline
  $f$ (\%)& $\tau$ (ps) & $\beta$ \\
    \hline
    5.6 & 1.6 $\pm$ 0.3 & 0.6 \\
    8.5 & 1.4 $\pm$ 0.3 & 0.6 \\
    10.5 & 1.6 $\pm$ 0.3 & 0.6 \\
    24.3 & 0.45 $\pm$ 0.10 & 0.6 \\
    \hline
  \end{tabular}
\end{table}
Moreover, it is impossible to extract a reliable value for stretching exponent; we then fixed it to the bulk value $\beta$= 0.6. The results collected in tab.~\ref{slowdata} show that the structural relaxation time does not vary appreciably on going from hydration $f$=5.6~\% to 10.5~\%, whereas the $f$=24.3~\% sample is characterized by a faster relaxation processes. The structural relaxation time of nanoconfined water at full hydration condition is indeed comparable with the bulk water time, $\tau$=0.35 ps. On the contrary, when the degree of hydration is less than two water layers, the structural relaxation experiences a clear slowing down, with an increase of the characteristic time scale of about a factor 4. These results suggest that the collective rearrangements of ``interfacial'' and ``inner'' water are characterized by similar structural processes (both require a stretched exponential function to be properly reproduced), but the characteristic time scales are clearly different\cite{gallo_00}.

\begin{figure}
\includegraphics[scale=.8]{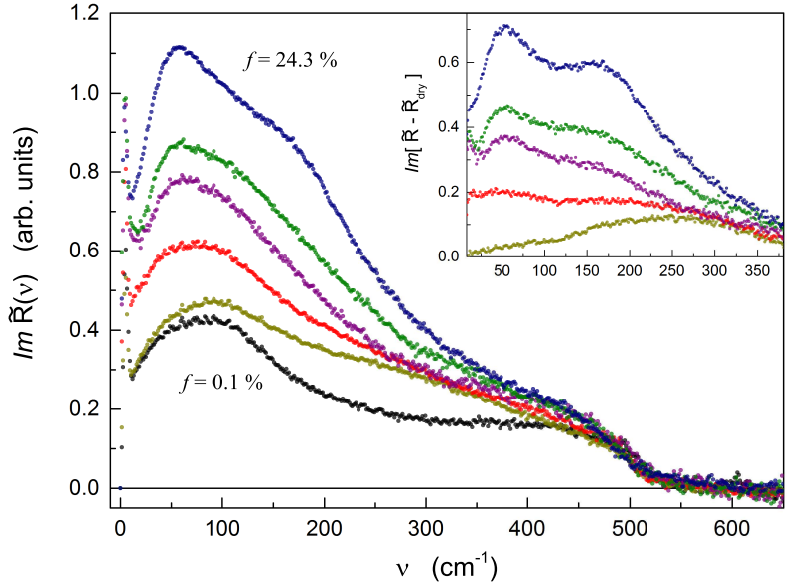}
\caption{
Fourier transforms of the HD-OKE data, deconvoluted from the instrumental response, at different filling fractions: from dry condition, $f$=0.1~\% to full hydration, $f$=24.3~\%. In the inset we report the same data after subtraction of the dry Vycor signal; The spectra give the contribution of nano-confined water.
}
\label{fig_4}
\end{figure}

In fig.~\ref{fig_4} we report the Fourier transform of the measured response function after deconvolution from the instrumental response, using the procedure summarized in the previous section and described in details in ref.\cite{taschin_13}. In the main panel of fig.~\ref{fig_4} we show the spectra, obtained by Fourier transformation of the HD/OKE data of fig.~\ref{fig_3}. The dry Vycor sample, $f$=0.1~\%, shows a frequency spectrum extending up to 500 cm$^{-1}$ characterized by a broad peak at about 100 cm$^{-1}$. We notice also: a narrow peak at about 5 cm$^{-1}$, hardly visible in the figure, whose nature was previously discussed in connection to the description of the time domain HD-OKE data, and a peak at 800 cm$^{-1}$, not shown in the figure. A part from the 5 cm$^{-1}$ band, the dry Vycor spectra reassemble closely the spectrum of amorphous SiO$_2$, as measured by depolarized Raman experiments\cite{pilla_03}. The presence of liquid water inside the Vycor pores causes an evident increase of the vibrational intensity in the frequency range from 20-350 cm$^{-1}$. In order to extract the contribution of nanoconfined water, we subtract from the data the dry Vycor spectrum, i.e. that for $f$=0.1~\%; the result is reported in the inset of fig.~\ref{fig_4}. At 3.3~\% hydration the spectrum of nanoconfined water is featureless; there are no signature of characteristic bending and stretching water peaks. At higher hydration, $f$=5.6~\%, the spectrum is unchanged in the high frequency part, $\nu \gtrsim$ 200 cm$^{-1}$, while on the low frequency side a weak signature of the bending peak appears. The spectra at the two highest levels of hydration show a growing presence of the bending and stretching features.
 
The present data suggest the following scenario for the dynamics of nanoconfined water.

Up to $f$=5.6~\%, corresponding to one water layer, only ``interfacial water'' exists inside the Vycor pores. This water is in direct contact with the surfaces, it is linked by the H-bonds formed with the silanol groups present on the surface. The water H-bond network of water is strongly modified by the surface interactions, these modifications affect the intermolecular vibrations spectrum. Our data show that this spectrum doesn't show any specific signature, suggesting a large network distortion characterized by a wide distribution of angles and distances. Moreover, the structural relaxation is much slower than in bulk water suggesting a strong hindering of the collective rearrangement and/or diffusion process.

Starting from $f$=10.5~\%, i.e. two water layers or pore plugs, part of the water molecules can form H-bonds with other water molecules without having a direct link to the surface silanol groups. In other words, we have some ``inner water''. This water present a liquid structure that reassemble the bulk water. The vibrational part of our data partially supports this point of view, even if the measured spectrum is still presenting several differences from that of bulk water. The measured inter-molecular vibrations indicate an H-bond network of nanoconfined water is characterized by deformations, which are larger and more numerous than in free bulk water. Also the structural phenomena are strongly affected by the surface interactions, as proved by the slowing down of the structural relaxation.  

At full hydration condition, $f$=24.3~\%, both the slow structural processes and the fast vibrational dynamics become similar to those of bulk water. This suggests that in this case the inner water shows several dynamic characteristics reminiscent the bulk liquid water.

\subsection{Comparison of bulk and nano-confined water dynamics}
\begin{figure}
\includegraphics[scale=.8]{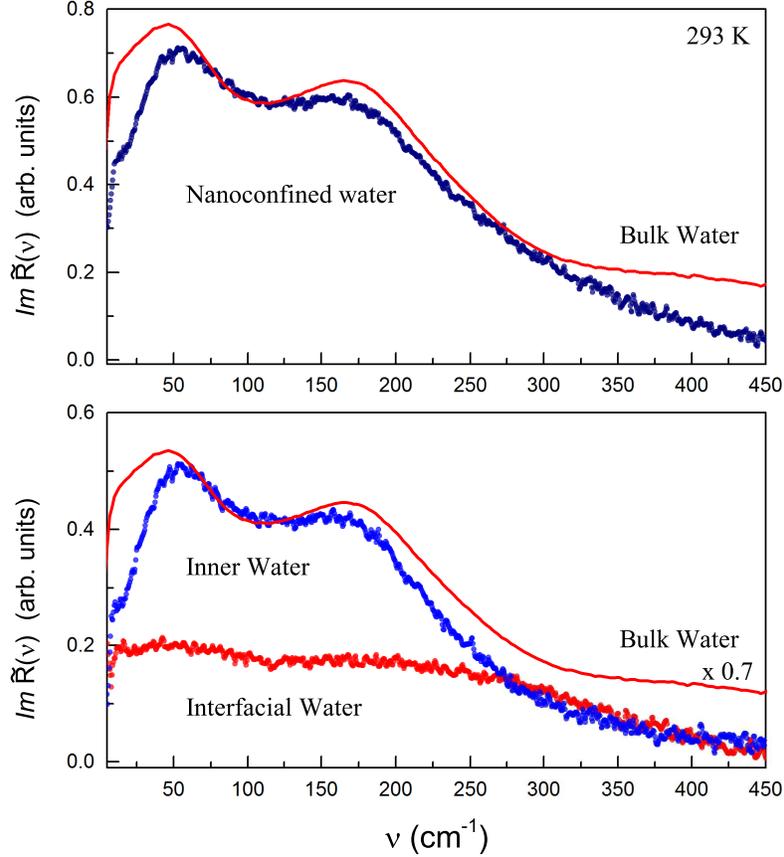}
\caption{Upper panel: comparison of the HD-OKE response function in the frequency of bulk water (red-line) and nanoconfined water (blue circles). Lower panel: decomposition of the total HD-OKE spectrum of nanoconfined water in the interfacial (red circles) and inner (blue circles) contributions.}
\label{fig_5}
\end{figure}

The direct comparison of nanoconfined and bulk water spectra, see upper panel of fig.~\ref{fig_5}, makes differences and similarities immediately evident. Clearly, both the spectra display the bending and stretching vibrational bands.

The low frequency part of the spectra, $\nu\leq$50 cm$^{-1}$, is mainly characterized by the large intensity difference of the band attributed to structural relaxation; this spectral component is much weaker in nanoconfined water, and this causes a steeper rise of the low frequency side of the bending band. At higher frequencies, the bending and stretching vibrational bands appear in both cases, but they are less defined in nanoconfined water then in bulk water. This is probably ascribable to a larger inhomogeneous broadening of those modes, due to the contribution of water in contact with the silica surfaces. At room temperature the water spectrum above 300 cm$^{-1}$ is generally attributed to “librational” modes\cite{desantis_04}; in nanoconfined water we observe a depletion of the librational components in this spectral region. In other words, it seems that nanoconfinement hinders the orientational oscillatory dynamics. 

Assuming that progressive hydration simply adds contributions to the HD-OKE signal, we applied a decomposition procedure in order to recover separately the contributions of interfacial and inner water dynamics.

The spectra of interfacial water is immediately obtained by subtracting the $f$=0.1~\% spectrum, assumed to represent that of the dry Vycor matrix, from the spectrum of the $f$=5.6~\% sample. The difference spectrum represents the contribution of the first water layer. By subtracting this interfacial water spectrum and that of the dry Vycor matrix from the spectrum measured for the fully hydrated sample ($f$=24.3~\%) we finally extracted the spectral contribution of the inner water. The results of this decomposition procedure are reported in the lower panel of fig.~\ref{fig_5}. The interfacial water shows a very board spectrum lacking the spectral features characteristic of the bulk phase. This unambiguous evidence proves that the H-bond network of interfacial water is substantially different from that of bulk water. Not surprisingly, the spectrum of inner water is largely similar to that of the bulk; in particular, the bending and stretching bands are present and well defined. The librational features of the spectrum above 300 cm$^{−1}$ are weakly present both in interfacial and inner water. In contrast, in the low frequency part ($\nu\leq$50 cm$^{−1}$) of the inner water spectrum we notice the same intensity reduction of the structural relaxation component observed in nanoconfined water.

As final conclusion, we can try to give an answer to the initial question asking to what extent the interaction with the pore surfaces modifies the water properties. According to our results for hydration levels $f\geq$10.5~\%, part of the nanoconfined water present structural and vibrational properties similar to those of bulk water. In other words, the liquid water added inside the pores above the $f$=10.5~\% amount, has large similarities with that characterizing the bulk water structure. At this level it is not possible to establish whether this is indicative of the presence of liquid plugs in the pores, or can be taken as evidence of the existence of real liquid polls. Nevertheless, we would like to stress out that the second water layer, even if not in direct contact with silica surface, has dynamic properties differing from those of bulk water. The van der Waals diameter of water molecule is of the order of 0.3 nm \cite{water_web_site}, so the thickness of the first two water layers covers about 1.2 nm of the pore diameter. If the pore diameter is 4 nm, as it is in Vycor porous glasses, there are about 4-5 layers of inner water; but if the pore diameter reduces below 1.5 nm there is no space available for inner water layer and all the water molecules are in the first or second layer. This suggests that when water is confined in a hydrophilic narrow pore, practically all the water molecules interact effectively with the surfaces, either directly (i.e first interfacial layer) or indirectly by water-water H-bond (i.e second layer). 

We believe that the findings reported in this work should be carefully taken into account when employing pore confinement in the attempt to approach experimentally the putative singularity temperature T$_s$ around 225 K in the supercooled phase of water. 

\section*{Acknowledgements}
The research has been performed at LENS. This work was supported by REGIONE TOSCANA POR-CRO-FSE 2007-2013 by EC COST Action MP0902-COINAPO. We acknowledge M. De Pas and A. Montori for providing their continuous assistance in the set-up of the electronics; R. Ballerini and A. Hajeb for the mechanical realizations; M.Pucci for the optical treatments of Vycor glasses

\end{document}